\documentclass[twocolumn,floatfix,showpacs]{revtex4}%
\usepackage{graphicx}%
\usepackage{amsmath}%
\usepackage{amsfonts}%
\usepackage{amssymb}
\usepackage{bm}

\def\s{{\sigma}}
\def\e{{\epsilon}}
\def\k{{ {\bm k} }}

\def\q{{ {\bm q} }}
\def\Q{{ {\bm Q} }}

\def\0{{ {\bm 0} }}
\def\w{{\omega}}
\def\a{{\alpha}}
\def\b{{\beta}}

\allowdisplaybreaks[4]

\begin{document}
\title{
Functional renormalization group study of 
orbital fluctuation mediated superconductivity:
Impact of the electron-boson coupling vertex corrections
}
\author{
Rina Tazai, 
Youichi Yamakawa, 
Masahisa Tsuchiizu, and
Hiroshi Kontani
}


\date{\today }

\begin{abstract}

In various multiorbital systems,
the emergence of the orbital fluctuations and
their role on the pairing mechanism 
attract increasing attention.
To achieve deep understanding on these issues,
we perform the functional-renormalization-group (fRG) study 
for the two-orbital Hubbard model.
The vertex corrections for the electron-boson coupling ($U$-VC),
which are dropped in the Migdal-Eliashberg gap equation,
are obtained by solving the RG equation.
We reveal that the dressed electron-boson coupling
for the charge-channel, ${\hat U}_{\rm eff}^c$,
becomes much larger than the bare Coulomb interaction, ${\hat U}^{0}$,
due to the $U$-VC in the presence of moderate spin fluctuations.
For this reason, the attractive pairing interaction due to 
the charge or orbital fluctuations is enlarged by the factor
$({\hat U}_{\rm eff}^c/{\hat U}^0)^2 \gg1$.
In contrast, the spin fluctuation pairing interaction
is suppressed by the spin-channel $U$-VC,
because of the relation ${\hat U}_{\rm eff}^s\ll{\hat U}^{0}$.
The present study demonstrates that the 
orbital or charge fluctuation pairing mechanism 
can be realized in various multiorbital systems thanks to the $U$-VC,
such as in Fe-based superconductors.

\end{abstract}

\address{
Department of Physics, Nagoya University,
Furo-cho, Nagoya 464-8602, Japan. 
}
 

\sloppy

\maketitle

\section{Introduction}
\label{sec:Intro}

Motivated by recent discoveries of interesting multiorbital superconductors,
unconventional pairing mechanisms 
driven by the orbital degrees of freedom
have attracted increasing attention.
For example, in FeSe families and some heavy fermion superconductors, 
the superconductivity (SC) appears next to the non-magnetic orbital order phase.
Such a phase diagram indicates a significant role of the 
orbital fluctuations on the pairing mechanism.

From a theoretical point of view,
it has been a big challenge to explain the emergence of the
orbital order/fluctuations based on realistic 
multiorbital Hubbard models microscopically.
In fact, only the spin fluctuations develop 
whereas the orbital fluctuations remain small
within the conventional mean-field-level approximations, 
such as the random-phase-approximation (RPA) and 
the fluctuation-exchange (FLEX) approximation
\cite{Bickers}.
Thus, non-magnetic orbital order cannot be explained
based on the mean-field-level approximations.
The reason for this failure would be that the 
interplay between orbital and spin fluctuations,
which is described by the vertex correction (VC),
is totally neglected in the RPA and FLEX.
Recently, the orbital order in Fe-based superconductors
has been naturally explained by taking the Aslamazov-Larkin VC (AL-VC)
into account \cite{Onari-SCVC,Onari-SCVCS,Yamakawa-FeSe}.

In order to study the VCs,
the functional-renormalization-group (fRG) 
is a very powerful and reliable theoretical method.
Both the charge-channel and spin-channel VCs are 
calculated in an unbiased way by solving the RG equation, 
since the particle-particle and particle-hole
channels are included on the same footing
without violating the Pauli principle.
Using the fRG theory, strong orbital fluctuation emerges
in two-orbital Hubbard models in the presence of moderate spin fluctuations, 
as revealed in Refs. \cite{Tsuchiizu1,Tsuchiizu2}.
These fRG studies confirmed the validity of the 
orbital fluctuation mechanism driven by the
orbital-spin mode-coupling due to the AL-VC
\cite{Onari-SCVC,Yamakawa-FeSe}.

Theoretically, it is natural to expect that 
the developed orbital fluctuations mediate the pairing formation.
The orbital fluctuations can induce not only the 
singlet SC (SSC), but also the triplet SC (TSC).
By performing the fRG theory for the multiorbital models for Sr$_2$RuO$_4$,
in which the TSC ($T_{\rm c}=1.5$ K) is expected to be realized
 \cite{Maeno,Maeno2,Sigrist-Rev,Ishida,Nomura,Wang,RG-Scaffidi,Kivelson},
orbital-fluctuation-mediated TSC has been proposed.
In the frequently-used Migdal-Eliashberg (ME) approximation,
the SSC pairing interaction is
$\frac32{\hat U}^{0;s}{\hat \chi}^s(q){\hat U}^{0;s}
-\frac12{\hat U}^{0;c}{\hat \chi}^c(q){\hat U}^{0;c}$,
and the TSC pairing interaction is
$-\frac12{\hat U}^{0;s}{\hat \chi}^s(q){\hat U}^{0;s}
-\frac12{\hat U}^{0;c}{\hat \chi}^c(q){\hat U}^{0;c}$,
where ${\hat U}^{0;c(s)}$ is the bare Coulomb interaction matrix
for the charge (spin) channel \cite{Onari-SCVC}.
Within the ME approximation,
spin-fluctuation-mediated SSC is expected 
when ${\hat \chi}^s(q)$ and ${\hat \chi}^c(q)$ are comparable,
because of the factor $\frac32$ for ${\hat \chi}^s(q)$
in the SSC pairing interaction.
However, this expectation is never guaranteed beyond the ME approximation
since ${\hat U}^{0;c}$ may be enlarged by the VC at low energies,
which is actually realized as we explain in the present paper.

In this paper,
we analyze the two-orbital Hubbard model
for the $(\a,\b)$-bands in Sr$_2$RuO$_4$ by using the fRG theory.
The aim of the present study is 
to confirm the realization condition for the 
orbital-fluctuation-mediated SC by going beyond the ME approximation.
For this purpose, we solve the gap equation by including 
the VC for the bare electron-boson coupling (EBC),
which we call the $U$-VC.
Due to the $U$-VC, the effective EBC for the charge (spin) channel,
${\hat U}^{c(s)}(k,k')$, deviates from 
the bare Coulomb interaction ${\hat U}^{0;c(s)}$.
By applying the fRG theory, we find the relation 
$|{\hat U}^{c}(k,k')|\gg |{\hat U}^{0;c}|$ due to the charge-channel $U$-VC
in the presence of moderate spin fluctuations.
In contrast, 
${\hat U}^{s}(k,k')$ is significantly suppressed by the spin channel $U$-VC
at low energies.
For these reasons, 
orbital-fluctuation-mediated SC will be realized 
in various multiorbital systems, 
such as in Fe-based superconductors and Sr$_2$RuO$_4$.
We stress that the phonon-mediated attractive pairing 
is also enlarged by the factor $({\hat U}^{c}(k,k')/{\hat U}^{0;c})^2$.

The Fermi liquid theory tells that the same $U$-VC causes 
(i) the enhancement of the orbital susceptibility and 
(ii) that of the orbital-fluctuation-mediated pairing interaction.
This fact means that (i) and (ii) are realized simultaneously.
This expectation will be confirmed by the present fRG study.

\section{$U$-VC for the susceptibilities and gap equation}
\label{sec:diagram}

\begin{figure}[!htb]
\includegraphics[width=.9\linewidth]{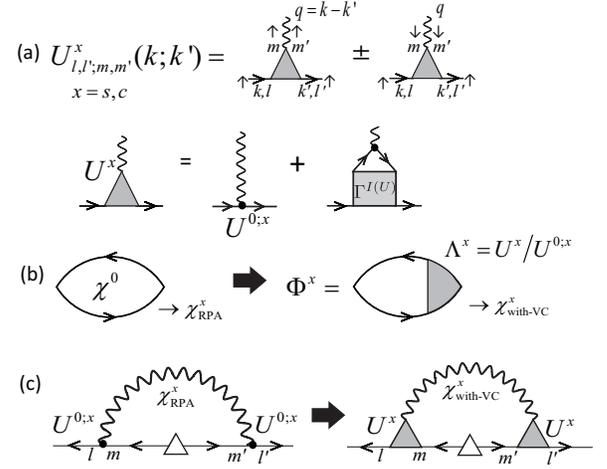}
\caption{
(a) The effective interaction ${\hat U}^{x}$ for $x=c$ ($+$) and $x=s$ ($-$),
which we call the dressed EBC.
The filled circle represents the Coulomb interaction ${\hat U}^{0;x}$,
and the rectangle ($\Gamma^{I(U),x}$) gives the $U$-VC. 
$\Gamma^{I(U),x}$ is irreducible with respect to ${\hat U}^{0;x}$
to avoid the double counting of the RPA-type diagrams.
(b) Beyond the RPA:
The irreducible susceptibility with the VC,
where ${\hat \Lambda}^{x}={\hat U}^{x}\{{\hat U}^{0;x}\}^{-1}$.
(c) Beyond the ME approximation:
The gap equation with the three-point VCs for the coupling constant ($U$-VC).
Only the single fluctuation exchange term is shown.
}
\label{fig:fig1}
\end{figure}

First, we introduce the dressed EBC due to the $U$-VC,
and formulate the susceptibilities ${\hat \chi}^{c,s}(q)$
and the gap equation in the presence of the same $U$-VC.
Figure \ref{fig:fig1} (a) shows the definition of the 
dressed EBC for the charge and spin channels,
${\hat U}^{c}(k,k')$ and ${\hat U}^{s}(k,k')$,
which are irreducible with respect to bare Coulomb interactions
${\hat U}^{0;c}$ and ${\hat U}^{0;s}$:
The definitions of ${\hat U}^{0;c}$ and ${\hat U}^{0;s}$
in the orbital basis are given in later section, 
and they were introduced in Refs. \cite{Takimoto,Onari-SCVC}.
We put $k=(\k,\e_n)=(\k,(2n+1)\pi T)$ and 
$q=(\q,\w_l)=(\q,2l\pi T)$ hereafter.
The solid and wavy lines represent the electron Green function ${\hat G}(k)$
and ${\hat \chi}^{x}(q)$ ($x=c,s$), respectively.
The rectangle ($\Gamma^{I(U),x}$) is the VC for the bare EBC ${\hat U}^{0;x}$,
which we call the $U$-VC. 
$\Gamma^{I(U),x}$ is irreducible with respect to ${\hat U}^{0;x}$
to avoid the double counting of the RPA-type diagrams.
In the present fRG study,
the $U$-VC is automatically obtained in solving the RG equation.
In later section, we also calculate $U$-VC 
due to the Aslamazov-Larkin term perturbatively,
which is the second-order terms with respect to ${\hat \chi}^{x}(q)$.

In Fig. \ref{fig:fig1} (b), we explain 
the VC for the irreducible susceptibility:
The bare susceptibility without the VC is 
$\chi_{l,l',m,m'}^0(q)= -T\sum_{n}G_{l,m}(k+q)G_{m',l'}(k)$,
where $G_{l,m}(k)$ is the Green function in the orbital basis.
Then, the RPA susceptibility is 
${\hat \chi}^x_{\rm RPA}(q)
={\hat \chi}^0(q)[{\hat 1}-{\hat U}^{0;x}{\hat \chi}^0(q)]^{-1}$.
By using the three-point vertex
${\hat \Lambda}^{x}={\hat U}^{x}\{{\hat U}^{0;x}\}^{-1}$,
the dressed irreducible susceptibility is given as
$\Phi^x(q)= -T\sum_{n}G(k+q)G(k)\Lambda^{x}(k+q,k)$,
where the orbital indices are omitted for simplicity.
Then, the susceptibility with full VCs is obtained as
${\hat \chi}^x_{\rm with \mbox{-} VC}(q)
={\hat \Phi}^x(q)[{\hat 1}-{\hat U}^{0;x}{\hat \Phi}^x(q)]^{-1}$.

Figure \ref{fig:fig1} (c) shows the gap equation 
due to the single-fluctuation-exchange term
in the presence of the $U$-VC for the EBC.
Within the RPA and the ME approximation,
the pairing interaction for the singlet state is
${\hat V}_{s,{\rm RPA}}(k,k')=\frac32 {\hat I}_{\rm RPA}^s(k-k')
-\frac12 {\hat I}_{\rm RPA}^c(k-k')-{\hat U}^{0;s}$,
where 
${\hat I}_{\rm RPA}^x(q)= {\hat U}^{0;x}
({\hat \chi}^x_{\rm RPA}(q)+\{{\hat U}^{0;x}\}^{-1}){\hat U}^{0;x}$. 
By including the VCs for both ${\hat \chi}^x_{\rm RPA}$ and 
the coupling constant ${\hat U}^{0;x}$, the 
pairing interaction with full VCs is given as
${\hat V}_{s,{\rm with\mbox{-}VC}}(k,k')=\frac32 {\hat I}_{\rm with\mbox{-}VC}^s(k,k')
-\frac12 {\hat I}_{\rm with\mbox{-}VC}^c(k,k')-{\hat U}^{0;s}$, 
where 
${\hat I}_{\rm with\mbox{-}VC}^x(k,k')= {\hat U}^{x}(k,k')
({\hat \chi}^x_{\rm with\mbox{-}VC}(k-k')+\{{\hat U}^{0;x}\}^{-1}){\hat U}^{x}(-k,-k')$. 

Therefore, the enhancement of the pairing interaction 
due to the charge-channel $U$-VC is naturally expected 
when the orbital fluctuations are realized by the $U$-VC,
in terms of the Fermi liquid theory.

For the purpose of analyzing the $U$-VC,
the fRG theory is very useful since the $U$-VC for 
${\hat \chi}^{x}(q)$ ($x=s,c$) and that for the gap equation
are generated on the same footings in terms of the parquet approximation.
This is a great merit of the fRG theory
\cite{RG-Review}.
In the present study, we use the RG+cRPA method,
which enables us to perform very accurate numerical study
\cite{Tsuchiizu1}.

\section{RG+cRPA study for the two-orbital Hubbard model}
\label{sec:RG-exp}
In this section, we analyze the 2-orbital 
($d_{xz}$, $d_{yz}$) Hubbard model,
as a canonical simple multiorbital systems.
We apply the 
renormalization-group plus constrained-RPA (RG+cRPA) method,
which was developed in Refs. \cite{Tsuchiizu1,Tsuchiizu2,Tsuchiizu3}.
By solving the RG differential equation, we obtain the
renormalized 4-point vertex $\hat{\Gamma}^{x}_{{\rm RG}}$ ($x=s,c$)
and susceptibilities $\chi ^{c(s)}(q)$ 
by taking account of the $U$-VC
in a systematic and in an unbiased way.
The superconducting state and the
transition temperature ($T_{\rm c}$)
are obtained by calculating the SSC and TSC 
susceptibilities, as formalized and performed 
in Ref. \cite{Tsuchiizu2}.

\subsection{Model Hamiltonian and the four-point vertex given by the RG+cRPA}
\label{sec:UVC1}

\begin{figure}[htb]
\includegraphics[width=.9\linewidth]{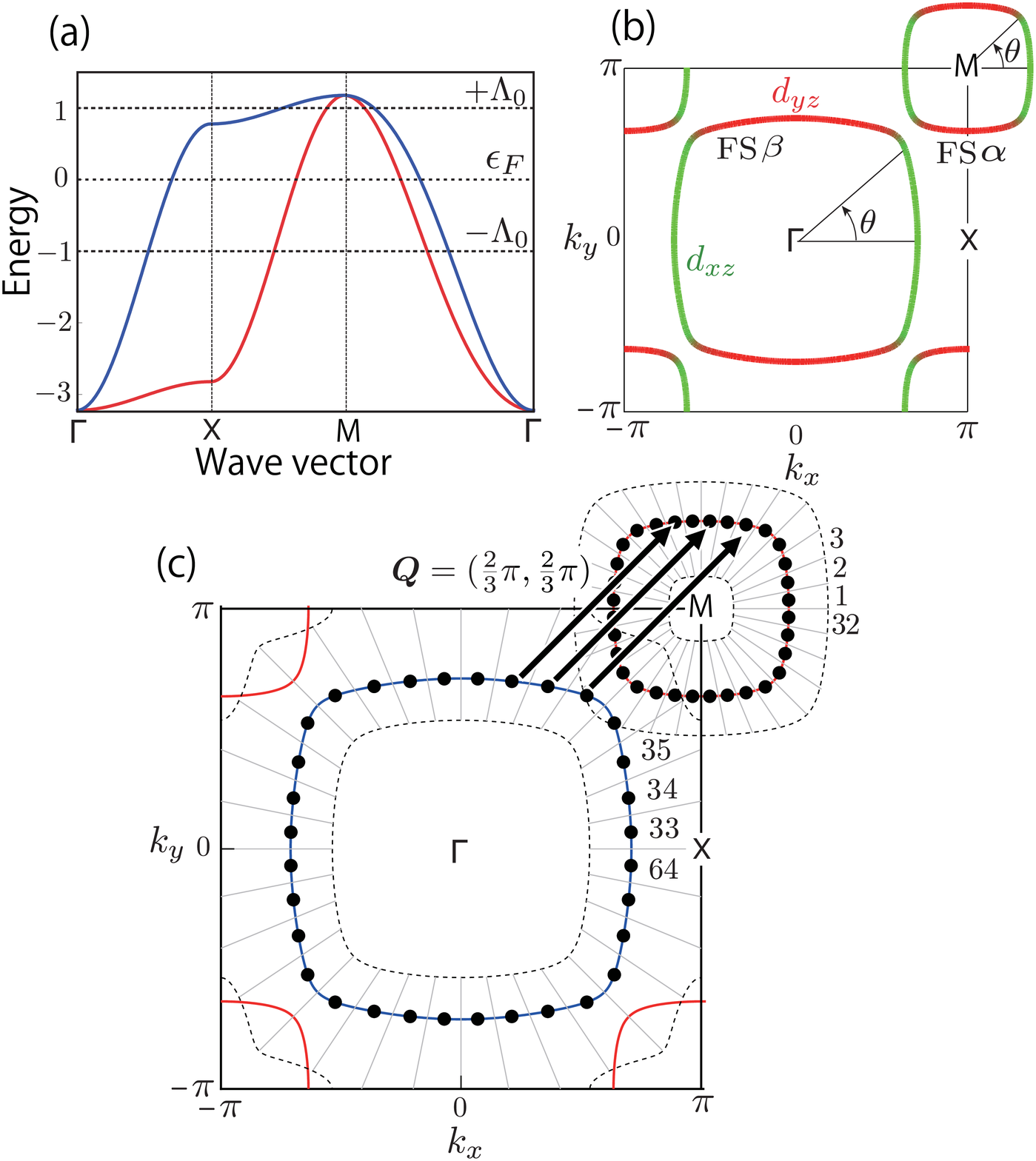}
\includegraphics[width=.9\linewidth]{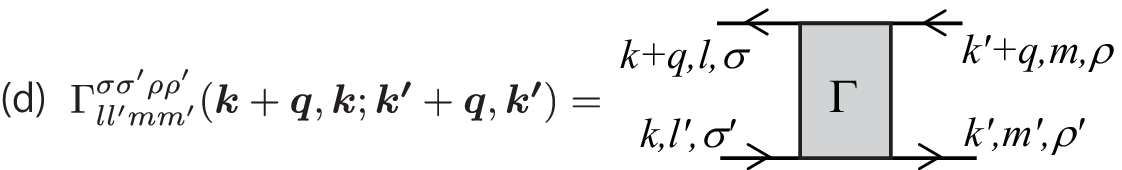}
\caption{(Color online)
(a) Band dispersion of 2-orbital Hubbard model 
and (b) FSs composed of the $d_{xz}$-orbital (green) 
and $d_{yz}$-orbital (red).
(c) The centre of patches ($1\sim64$) on the FSs.
The arrows represents the nesting vector.
The tip and the tail of each arrow correspond to
$(i_\a,i_\b)=(6,37),(8,38),(10,39)$.
(d) Definition of the full four-point vertex
$\Gamma_{ll'mm'}^{\sigma \sigma' \rho \rho'}(\k+\q,\k;\k'+\q,\k')$
in the microscopic Fermi liquid theory.
}
\label{fig:FS}
\end{figure}

First, we introduce the 2-orbitals square lattice Hubbard model,
which describes the ($d_{xz}$, $d_{yz}$)-orbital bandstructure
in $\rm{Sr_{2}RuO_{4}}$.
We set the kinetic term of the Hamiltonian as
\begin{eqnarray}
H_{0}=\sum_{k,\sigma}\sum_{l,m}\xi_{k}^{l,m}
c^{\dagger}_{k,l,\sigma}c_{k,m,\sigma} ,
\label{eqn:H0}
\end{eqnarray}
where $l, m$ takes $1$ or $2$, which corresponds to $d_{xz}$ or $d_{yz}$. 
$\xi^{l,m}_{k}$ is defined as 
$\xi^{1,1}_{k}=-2t\cos k_{x} -2t^{''}\cos k_{y}$, 
$\xi^{2,2}_{k}=-2t\cos k_{y} -2t^{''}\cos k_{x}$, 
$\xi^{1,2}_{k}=\xi^{2,1}_{k}=-4t^{'}\sin k_{x} \sin k_{y}$. 
Hereafter, we set the hopping parameters 
($t^{}$, $ t^{'}$, $ t^{''})=(1, 0.1, 0.1)$:
The unit of energy in the present study is $t=1$.
The number of electrons is fixed as $n=n_{xz}+n_{yz}=4\times (2/3)=2.67$. 
The obtained band dispersion and Fermi surfaces (FSs) are shown in 
Figs. \ref{fig:FS} (a) and (b), which reproduce 
FS{$\alpha$} and FS{$\beta$} in Sr$_2$RuO$_4$.
This model has been analyzed as a canonical multiorbital model 
in various theoretical studies,
such as the anomalous Hall effect
\cite{Kontani-AHE}.

In the RG+cRPA method, each band is divided into 
the higher-energy part ($|\e_{u,\k}|>\Lambda_0$) and
the lower-energy part ($|\e_{u,\k}|<\Lambda_0$).
In order to perform the renormalization procedure,
the lower-energy part is divided into $N_p/2$ patches.
Figure \ref{fig:FS} (c) shows the contours for 
$|\e_{u,\k}|=\Lambda_0=1$ and the center of patches $1\sim64$.

In addition, we introduce the on-site Coulomb interaction term,
which contains the intra-orbital and inter-orbital Coulomb interactions
$U$ and  $U'$, the Hund's coupling $J$, and the pair hopping interaction $J'$.
The bare Coulomb interaction term is expressed as
\begin{eqnarray}
&&\!\!\!\!\!\!\!\!\!\!\!\!
 H_{int}=\frac{1}{4}\sum_{i}\sum_{l l' m m'}
\sum_{\sigma \sigma' \rho \rho'}U_{ll'mm'}^{0;\sigma \sigma' \rho \rho'}
c^{\dagger}_{i l \sigma}
c_{i l' \sigma'}
c_{i m \rho} 
c^{\dagger}_{i m' \rho'} ,
\label{eqn:HUU} \\
&&\!\!\!\!\!\!\!\!\!\!\!\!
 U_{ll'mm'}^{0;\sigma \sigma' \rho \rho'}
=\frac{1}{2}U^{0;s}_{ll'mm'}
\vec{\bf{\sigma}}_{\sigma \sigma'} \cdot \vec{\bf{\sigma}}_{\rho' \rho}
+\frac{1}{2}U^{0;c}_{ll'mm'}\delta_{\sigma,\sigma'}\delta_{\rho',\rho} ,
\label{eqn:HU}
\end{eqnarray}
where $U^{0;c}_{ll'mm'}=(-U, U'-2J, -2U'+J, -J', 0)$ and 
$U^{0;s}_{ll'mm'}=(U, U', J, J', 0)$  in the cases of 
($l=l'=m=m'$, $l=m\neq l'=m'$, $l=l'\neq m=m'$, $l=m'\neq l'=m$ and otherwise).
Hereafter, we assume the relation $J=J'=(U-U')/2$.

The antisymmetrized full four-point vertex
${\hat \Gamma}(\k+\q,\k;\k'+\q,\k')$,
which is the dressed vertex of the
bare vertex ${\hat U}^{0}$ in Eq. (\ref{eqn:HU})
in the microscopic Fermi liquid theory \cite{AGD},
is depicted in Fig. \ref{fig:FS} (d).
Reflecting the SU(2) symmetry of the present model,
${\hat \Gamma}$ is uniquely decomposed into the  
spin-channel and charge-channel four-point vertices
by using the following relation:
\begin{eqnarray}
&&\Gamma_{ll'mm'}^{\sigma \sigma' \rho \rho'}(\k+\q,\k;\k'+\q,\k')
\nonumber \\
&& \ \ \ \ \ \ =\frac{1}{2}\Gamma_{ll'mm'}^{s}(\k+\q,\k;\k'+\q,\k')
\vec{\bf{\sigma}}_{\sigma \sigma'} \cdot \vec{\bf{\sigma}}_{\rho' \rho}
\nonumber \\
&& \ \ \ \ \ \ \ 
+\frac{1}{2}\Gamma^{c}_{ll'mm'}(\k+\q,\k;\k'+\q,\k')
\delta_{\sigma,\sigma'}\delta_{\rho',\rho} ,
\label{eqn:Gamma2}
\end{eqnarray}
where $\sigma, \sigma', \rho, \rho'$ are spin indices.
We stress that ${\hat \Gamma}^{c,s}$ 
are fully antisymmetrized, so the requirement by 
the Pauli principle is satisfied.
We note that 
${\hat \Gamma}^{\uparrow\uparrow\uparrow\uparrow}
=\frac12 {\hat \Gamma}^c+\frac12 {\hat \Gamma}^s$,
${\hat \Gamma}^{\uparrow\uparrow\downarrow\downarrow}
=\frac12 {\hat \Gamma}^c-\frac12 {\hat \Gamma}^s$,
and 
${\hat \Gamma}^{\uparrow\downarrow\uparrow\downarrow}
={\hat \Gamma}^s$.

\subsection{RG+cRPA Theory}
\label{sec:RG+cRPA}

We analyze the present model by using the RG+cRPA method,
which was introduced in our previous papers 
\cite{Tsuchiizu1,Tsuchiizu2,Tsuchiizu3}
in detail.
In this method, we introduce the original cutoff energy $\Lambda_{0}$
in order to divide each band 
into the higher and the lower energy regions:
The higher-energy scattering processes are calculated by using the cRPA:
The lower-energy scattering processes are analyzed by 
solving the RG equation,
in which the initial vertices in the differential equation
are given by the cRPA.
The lower energy region is divided into $N_p/2$ patches 
for each band as shown in Fig. \ref{fig:FS} (c).

\begin{figure}[htb]
\includegraphics[width=.9\linewidth]{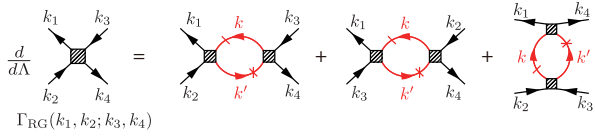}
\caption{(Color online)
The one-loop RG equation for the four-point vertex.
The crossed lines represent the electron Green function with cutoff 
$\Lambda$.
The slashed lines represent the 
electron propagations having the energy shell $\Lambda$.
}
\label{fig:FS2}
\end{figure}

In the RG formalism,
the four-point vertex function is determined 
 by solving the differential equations, called 
the RG equations.
In the band representation basis,
the explicit form of the RG equations is given by
\begin{widetext}
\begin{eqnarray}
\frac{d}{d\Lambda}
\Gamma_\mathrm{RG}(k_1,k_2;k_3,k_4)
 &=& 
-\frac{T}{N}\sum_{k,k'}
\left[
\frac{d}{d\Lambda}
 G(k) \, G(k')
\right]
\Bigl[
\Gamma_\mathrm{RG}(k_1,k_2;k,k') \, \Gamma_\mathrm{RG}(k,k';k_3,k_4)
\nonumber \\
&& {}
- \Gamma_\mathrm{RG}(k_1,k_3;k,k') \, \Gamma_\mathrm{RG}(k,k';k_2,k_4) 
- \frac{1}{2} \Gamma_\mathrm{RG}(k_1,k; k',k_4) \, 
  \Gamma_\mathrm{RG}(k,k_2;k_3,k')
\Bigr] ,
\end{eqnarray}
\end{widetext}
where $G(k)$ is the Green function multiplied by the 
Heaviside step function $\theta(|\e_{u,\k}|-\Lambda)$, and 
$k$ is the compact notation of the momentum, band, and spin index:
$k=(\k, \e_n, u, \sigma)$.
The diagrammatic representation of the RG equations 
is shown in Fig.\ \ref{fig:FS2}.
The first two contributions in the rhs represent the particle-hole
  channels and the last contribution is the 
 particle-particle channel.

The four-point vertex $\Gamma_\mathrm{RG}(k_1,k_2;k_3,k_4)$
is obtained by solving the above RG  differential equation 
from $\Lambda_0$ to the lower cutoff energy $\w_c$.
In a conventional fRG method,
$\Lambda_0$ is set larger than the bandwidth $W_{\rm band}$,
and the initial value is given by the bare Coulomb interaction 
in Eq. (\ref{eqn:HU}).
In the RG+cRPA method, 
we set $\Lambda_0<W_{\rm band}$, and the initial value is given by
the constraint RPA to include the higher-energy processes
without over-counting of diagrams \cite{Tsuchiizu1}.

The merits of the RG+cRPA method are listed as:
(i) The higher-energy processes are accurately calculated 
within the cRPA by introducing the fine (such as $128\times128$) $\k$-meshes.
This method is justified since the VCs are less important at higher energies.
In the conventional $N_p$-patch fRG method,
numerical errors due to the violation of the momentum-conservation 
becomes serious at higher-energy processes.
(ii) The scattering processes contributed by the valence-bands
(=Van-Vleck processes), which are important in multiorbital systems
to derive physical orbital susceptibility,
are taken into account in the RG+cRPA method.
Especially, the Van-Vleck processes are crucial to 
obtain the orbital susceptibilities without unphysical behaviors.

The full four-point vertex in Fig. \ref{fig:FS} (d)
is expressed in the band basis.
On the other hand, we solve the four-point vertex
in the orbital basis in the present RG+cRPA study, expressed as
${\Gamma}_{uu'vv'}^{\s\s'\rho\rho'}(\k_1,\k_2;\k_3,\k_4)$.
These expressions are transformed to each other
by using the unitary matrix $u_{l,u}(\k)=\langle l,\k|u,\k \rangle$.
In the present RG+cRPA study, 
we assume that each $\k_i$ is on the FSs, so
we are allowed to drop four band indices $u,u',v,v'$.

In this paper, we set $\Lambda_{0}=1.0$ ($<$ band width) and $N_p=64$,
and introduce the logarithmic energy scaling parameter 
$\Lambda_{l}=\Lambda_{0}e^{-l}$ ($l\ge0$)
in solving the RG equation.
We verified that reliable results are obtained 
by setting $\Lambda_{0}\sim W_{\rm band}/2$.

\subsection{Phase diagram obtained by the RG+cRPA}
\label{sec:RG}

\begin{figure}[htb]
\includegraphics[width=.9\linewidth]{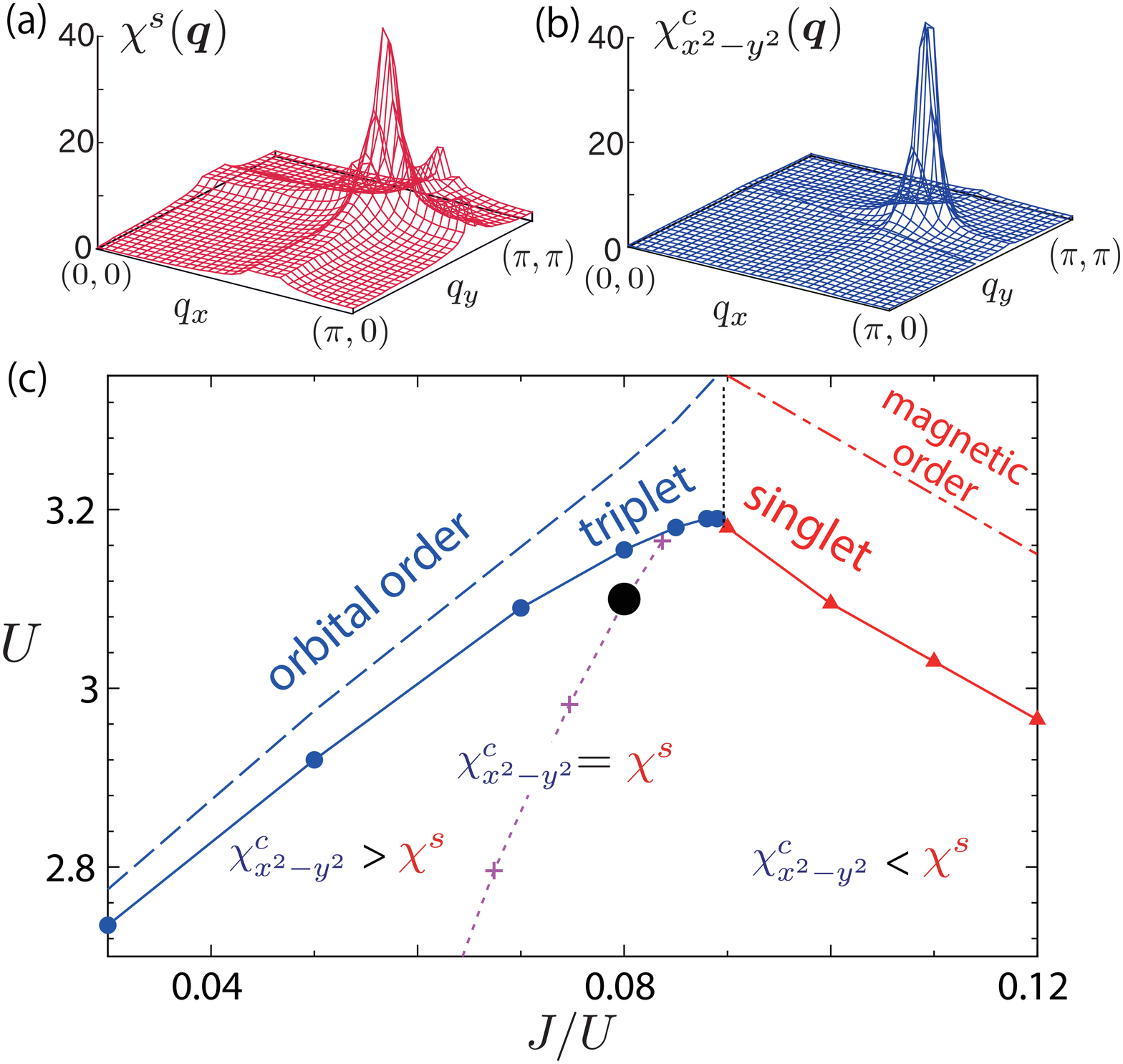}
\caption{(Color online)
(a) $\q$-dependence of obtained total spin susceptibility $\chi^{s}(\q)$ 
enlarged at ${\bm q} \approx (2\pi/3,2\pi/3)$. 
(b) Obtained quadrupole susceptibility $\chi^c_{x^2-y^2}(\q)$. 
(c) SC phase diagram obtained by RG+cRPA method.
}
\label{fig:phase}
\end{figure}

First, we calculate the spin/charge susceptibilities
and SSC/TSC susceptibilities at $T=5\times 10^{-4}$ 
by performing the RG+cRPA analysis.
The renormalization is fulfilled till $\Lambda_{l}$
reaches $\Lambda_{l_c}=10^{-2}T$
(i.e., $l_c={\rm ln}(\Lambda_0/10^{-2}T)$).
The charge (spin) susceptibilities in the multiorbital model is
\begin{eqnarray}
\chi^{c(s)}_{l l' m m'}(q)=\int^{\beta}_{0} d\tau \frac{1}{2}\left\langle A^{c(s)}_{l l'}({\bm q},\tau)A^{c(s)}_{m' m}({\bm -\q},0)\right\rangle e^{i\w_l\tau},
\label{eq:suscep}
\end{eqnarray}
where
\begin{eqnarray}
A^{c(s)}_{l\, l'}({\bm q})=\sum_{\bm k}
(c^{\dagger}_{{\k} l' \uparrow}
c_{{\k+\q} l \uparrow}+(-)
c^{\dagger}_{{\bm k} l' \downarrow}
c_{{\k+\q} l \downarrow}) .
\label{eqn:A}
\end{eqnarray}
The obtained susceptibilities are shown in the 
Figs. \ref{fig:phase} (a) and (b) :
$\chi^c_{x^2-y^2}({\bm q})=\sum_{l , m}(-1)^{l+m}\chi^{c}_{l, l, m, m}({\bm q})$ 
is the orbital susceptibility with respect to the 
orbital polarization $n_{xz}-n_{yz}$,
and $\chi^{s}({\bm q})=\sum_{l , m}\chi^{s}_{l, l, m, m}({\bm q})$
is the total spin susceptibility.
We set the parameters $(U, J/U)=(3.10, 0.08)$ and $T=5\times10^{-4}$,
which corresponds to the black circle in 
the phase diagram in Fig. \ref{fig:phase} (c).
Both $\chi^{s}(\q)$ and $\chi^c_{x^2-y^2}(\q)$ 
has the maximum around the nesting vector ${\bm Q}=(2\pi/3, 2\pi/3)$,
and the relation $\chi^{s}(\Q)\approx \chi^c_{x^2-y^2}(\Q)$ is realized. 
The strong peak in $\chi^{s}(\Q)$ has been observed by the 
neutron inelastic scattering study for Sr$_2$RuO$_4$
 \cite{neutron}.
In addition to this result, the STM study \cite{STM}
indicates that the TSC in Sr$_2$RuO$_4$
mainly originates from the electronic correlation in the ($\a,\b$)-bands.
We stress that the strong enhancement of $\chi^c_{x^2-y^2}$ 
cannot be obtained in the RPA.
This fact means that the strong orbital fluctuations
originate from the $U$-VC, shown in Fig. \ref{fig:fig1} (b),
calculated by the RG method appropriately.

Secondly, we calculate the TSC (SSC) susceptibilities 
$\chi^{\rm{SC}}_{t(s)}$ by the RG+cRPA method. 
It is defined as    
\begin{eqnarray}
\chi^{\rm{SC}}_{t(s)}= \frac{1}{2}\int^{\beta}_{0} d\tau
\left\langle B^{\dagger}_{t(s)}(\tau)B_{t(s)}(0)\right\rangle ,
\label{eqn:suscepSC}
\end{eqnarray}
where
\begin{eqnarray}
B_{t(s)}=\sum_{{\bm k}\in {\rm FS}}\Delta_{t(s)}({\bm k})
c_{{\bm k},\uparrow}c_{{\bm -\k},\uparrow(\downarrow)} .
\label{eqn:B}
\end{eqnarray}
%
The gap function $\Delta_{t(s)}({\bm q})$ 
in Eq. (\ref{eqn:B}) is uniquely determined 
by maximizing the SC susceptibilities 
\cite{Tsuchiizu2}.

The obtained numerical results for 
$T=5\times10^{-4}$ and $\Lambda_{l_c}=10^{-2}T$
are summarized as the phase diagram in Fig. \ref{fig:phase} (c).
The boundary of the orbital and magnetic orders are shown 
by the broken lines, and the relation 
$\chi^s(\Q)= \chi^c_{x^2-y^2}(\Q)$ holds on the dotted line.
The boundaries for the TSC and SSC transition
are shown by the solid lines.
Thus, the TSC and SSC states are respectively realized below the 
orbital and magnetic order boundaries,
for wide range of parameters.
We stress that the strong orbital fluctuations and the TSC state
is obtained for $J/U\lesssim O(0.1)$,
which is comparable to the ratio $J/U=0.0945$ in FeSe
derived from the first-principles study.
The present result is substantially improved 
compared to the previous phase diagram for $\Lambda_0=1$
in Ref. \cite{Tsuchiizu2},
in which the strong orbital fluctuations appear only for $J/U<0.03$.
The reason for this improvement is that 
four-point vertex in Ref. \cite{Tsuchiizu2} is underestimated
since we included only the processes that rigorously satisfy 
the momentum conservation in solving the RG equation.
In the present study, we allow the scattering processes
if the momentum conservation is satisfied within the patch resolution,
according to a similar manner explained in 
Ref. \cite{Metzner,Honerkamp,RG-Review}.
This improved method was utilized in the study of the 
charge-density-wave 
in curate superconductors \cite{Tsuchiizu3}.

The obtained TSC gap function belongs to the $E_{u}$ representation,
and approximately follows the following $\k$-dependence:
($\Delta_{t,x}({\bm k})$,$\Delta_{t,y}({\bm k})$)
$\propto (\sin 3k_{x},\sin 3k_{y})$.
The SSC gap function belongs to $A_{1g}$ or $B_{1g}$ symmetry
in the phase diagram in Fig. \ref{fig:phase} (c),
similarly to our previous study in Ref. \cite{Tsuchiizu2}.

Until now, many theoretical studies on the mechanism of the 
TSC in Sr$_2$RuO$_4$ have been performed.
They are roughly classified into the following two scenarios.
One of them is that the TSC is realized mainly 
in a two-dimensional (2D) FS$\gamma$ composed by the $d_{xy}$-orbital \cite{Nomura, Wang}.
Nomura and Yamada explained the TSC state by using 
the higher-order perturbation theory \cite{Nomura}.
In addition, Wang {\it et al}. performed the 2D RG and
discussed that the TSC is realized on the FS$\gamma$  in the presence of spin fluctuations at 
$\q =(0.19\pi,0.19\pi)$. On the other hand, the TSC originating from the q1D FSs had been discussed
 by applying the perturbation theory \cite{Kivelson, RG-Scaffidi} and the RPA \cite{Takimoto}.
Takimoto proposed the orbital-fluctuation-mediated TSC  in the RPA \cite{Takimoto}. However, 
under the realistic condition $U' < U$, the TSC could not overwhelm the SSC in the RPA. 
In contrast to the RPA,  the present authors obtained the TSC state in the wide parameters range with 
realistic condition $U' < U$ by using the RG+cRPA theory. 
As shown in the following section, these results originate from the important roles 
of the $U$-VC which is neglected in the RPA. 

From the experimental aspect, many efforts have been devoted to reveal the electronic state and the gap structure in Sr$_2$RuO$_4$.
For example, strong AFM fluctuations at $\Q$ by the nesting of $\alpha$ and $\beta$ bands were observed by neutron scattering spectroscopy \cite{neutron}.
In addition, a large SC gap with 2$|\Delta| \approx 5T_{c}$ was observed by the scanning tunneling microscopy measurement \cite{STM}.  The authors expected that the observed large gap appears on the q1D FSs, since the tunneling will be dominated by the ($d_{xz}$,$d_{yz}$) orbitals that stand along the $z$ axis.  These experiments indicate that the active bands of the TSC in Sr$_2$RuO$_4$ is q1D FSs.

\section{Origin of orbital fluctuation mediated SC: 
Significant Role of the $U$-VC}

In the previous section, we explained that the
orbital-fluctuation-mediated TSC state is obtained
for realistic parameter range by using the improved RG+cRPA method.
In this section, we reveal the microscopic reason
why the orbital-fluctuation-mediated pairing interaction
becomes superior to the spin-fluctuation-mediated one 
in the case that ${\hat \chi}^s(q)$ and ${\hat \chi}^c(q)$ are comparable.
This is the main aim of the present paper.

\subsection{Gap equation beyond the ME scheme
}

Here, we study the SC state by analyzing the linearized gap equation
based on the pairing interaction obtained by the RG equation
\cite{RG-gapeq}.
The gap equation in the band basis is given as
\begin{eqnarray}
&&\lambda_{t(s)} \Delta_{t(s)}({\bm k}) =\nonumber \\
&&-\int_{\rm FS} \frac{d{\bm k'}}{v_{{\bm k'}}} {V}^{\w_c}_{t(s)}({\bm k},{\bm k'}) \Delta_{t(s)}({\bm k'}) \ln{\frac{1.13\omega_{c}}{T}} ,
  \label{eqn:gap-eq}
\end{eqnarray}
where $\Delta_{t(s)}({\bm k})$ is the TSC (SSC) gap function on the FSs, 
which has odd (even) parity.
In Eq. (\ref{eqn:gap-eq}),
$\k$ and $\k'$ are the momenta on the FS$\a$ and FS$\beta$,
$\lambda_{t(s)}$ is the eigenvalue of the gap equation,
and ${V}^{\w_c}_{t(s)}$ is the pairing interaction
given by the RG equation, by setting the lower-energy cutoff
as $\Lambda_{l_c}= \omega_c$ (i.e., $l_c= {\rm ln}(\Lambda_0/\omega_c)$).
The expression of the pairing interaction is given below.
We choose the cutoff $\omega_c$ so as to satisfy $\w_c \gg T$,
and assume that the renormalization of the susceptibilities
${\hat \chi}^{s,c}(\q)$ saturates for $\Lambda_l<\w_c$.
In deriving Eq. (\ref{eqn:gap-eq}), we used the relation
$\int_{-\w_c}^{\w_c} d\e_{\k'} \frac1{2\e_{\k'}}{\rm th}(\e_{\k'}/2T)
= {\rm ln} (1.13\w_c/T)$.


In the present RG study,
the pairing interaction in the band 
is directly given by solving the RG equation 
for the four-point vertex ${\Gamma}_{{\rm RG}}$,
till the lower-energy cutoff $\Lambda_{l_c}= \omega_c$.
We set $\w_c=12T= 6\times 10^{-3}$.

\begin{figure}[htb]
\includegraphics[width=.8\linewidth]{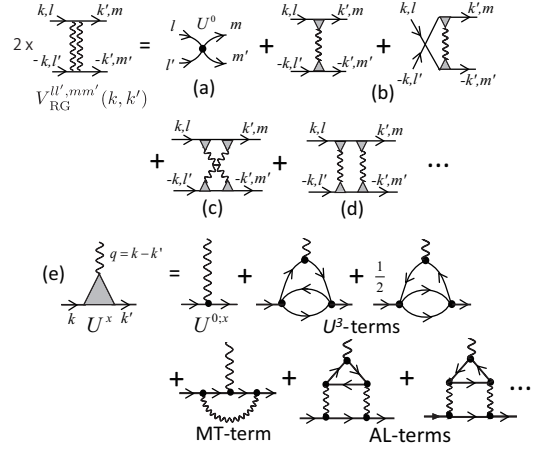}
\caption{
(a) The bare interaction, (b) single-fluctuation-exchange term,
(c) crossing-fluctuation-exchange term, and 
(d) the lowest particle-particle term.
(e) Typical diagrams for the $U$-VC.
For the charge sector,
the Maki-Thompson (MT) term is negligibly smaller than the AL term
in the presence of moderate spin fluctuations.
The $O(\{U^0\}^3)$-terms in MT and AL terms are dropped
to avoid the double counting.
In (a)-(e), spin indices are not  written explicitly.}
\label{fig:diagram}
\end{figure}

By using the four-point vertex given by the RG+cRPA
in the band basis representation,
the pairing interaction in Eq. (\ref{eqn:gap-eq})
with the $U$-VC is given as
\begin{eqnarray}{V}^{}_{t,{\rm RG}}({\bm{k},\bm{k}'})&=&
-\frac{1}{4}{\Gamma}^{s}_{{\rm RG}}(\k,\k';-\k',-\k)
 \nonumber \\
&&-\frac{1}{4}{\Gamma}^{c}_{{\rm RG}}(\k,\k';-\k',-\k) ,
 \label{eqn:V1t} \\
{V}^{}_{s,\rm{RG}}(\k,\k')&=&
\frac{3}{4}{\Gamma}^{s}_{{\rm RG}}(\k,\k';-\k',-\k)
 \nonumber \\
&&-\frac{1}{4}{\Gamma}^{c}_{{\rm RG}}(\k,\k';-\k',-\k) .
\label{eqn:V1s}
\end{eqnarray}
In ${V}^{}_{t(s),{\rm RG}}(\k,\k')$,
the $U$-VC for the pairing interaction shown in Fig. \ref{fig:fig1} (c)
is automatically included.
In Fig. \ref{fig:diagram},
we show the typical diagrams included in ${\Gamma}_{\rm RG}$:
The bare Coulomb interaction term is given in Fig. \ref{fig:diagram} (a).
The single- and crossing-fluctuation-exchange terms are 
shown in Figs. \ref{fig:diagram} (b) and (c), respectively.
The particle-particle ladder term is shown in Fig. \ref{fig:diagram} (d),
which is expected to be small when $\w_c\gg T_{\rm c}$.
The typical diagrams for the $U$-VC are shown in Fig. \ref{fig:diagram} (e).

In order to verify the importance of the $U$-VC, 
we also introduce the pairing interaction within the ME scheme:
For this purpose, we solve the RG equation for ${\hat \chi}^{c(s)}_{\rm RG}$
till the lower cutoff $\Lambda_{l_c}=\w_c$.
We set $\w_c=12T= 6\times 10^{-3}$.
Using the obtained ${\hat \chi}^{c(s)}_{\rm RG}$,
the antisymmetrized four-point vertex
in the single-fluctuation-exchange approximation is expressed
in the orbital basis as follows:
\begin{eqnarray}
&&{\Gamma}^{s}_{\chi,12,34}= \hat{U}^{0;s}_{12,34}
+(\hat{U}^{0;s}\hat{\chi}^{s}(1-2)\hat{U}^{0;s})_{12,34}
\nonumber \\
&&\ \ \ \ \ \ \ \ \ 
-\frac{1}{2}(\hat{U}^{0;c}\hat{\chi}^{c}(1-3)\hat{U}^{0;c})_{13,24}
\nonumber \\
&&\ \ \ \ \ \ \ \ \ 
+\frac{1}{2}(\hat{U}^{0;s}\hat{\chi}^{s}(1-3)\hat{U}^{0;s})_{13,24} ,
 \label{eqn:V3s} \\
&&{\Gamma}^{c}_{\chi,12,34}=\hat{U}^{0;c}_{12,34}
+(\hat{U}^{0;c}\hat{\chi}^{c}(1-2)\hat{U}^{0;c})_{12,34}
\nonumber \\
&&\ \ \ \ \ \ \ \ \ 
-\frac{1}{2}(\hat{U}^{0;c}\hat{\chi}^{c}(1-3)\hat{U}^{0;c})_{13,24}
\nonumber \\
&&\ \ \ \ \ \ \ \ \ 
-\frac{3}{2}(\hat{U}^{0;s}\hat{\chi}^{s}(1-3)\hat{U}^{0;s})_{13,24} .
\label{eqn:V3c}
\end{eqnarray}
Here, $\hat{U}^{0;c(s)}$ is the bare Coulomb interaction
in Eq. (\ref{eqn:HU}), and
$\hat{\chi}^{c(s)}_{{\rm RG}}$ is the $(2\times2)\times(2\times2)$ matrix.
The diagrammatic expression for $\hat{V}^{}_{t(s),\chi}$
is given by dropping the $U$-VC in Fig. \ref{fig:diagram} (b).

The pairing interaction $V_{t,\chi}(\k,\k')$ [$V_{s,\chi}(\k,\k')$]
in the absence of the $U$-VCs are obtained by 
inputting Eqs. (\ref{eqn:V3s})-(\ref{eqn:V3c})
into Eq. (\ref{eqn:V1t}) [Eq. (\ref{eqn:V1s})], respectively,
after performing the unitary transformation by using $u_{l,u}(\k)$.
Then, ${\hat \chi}^{s,c}(1-2)$ [${\hat \chi}^{s,c}(1-3)$] 
in Eqs. (\ref{eqn:V3s}) and (\ref{eqn:V3c}) is replaced with 
${\hat \chi}^{s,c}(\k-\k')$ [${\hat \chi}^{s,c}(\k+\k')$].

\subsection{Analysis of the $U$-VC based on the RG+cRPA method}

\begin{figure}[htb]
\includegraphics[width=.9\linewidth]{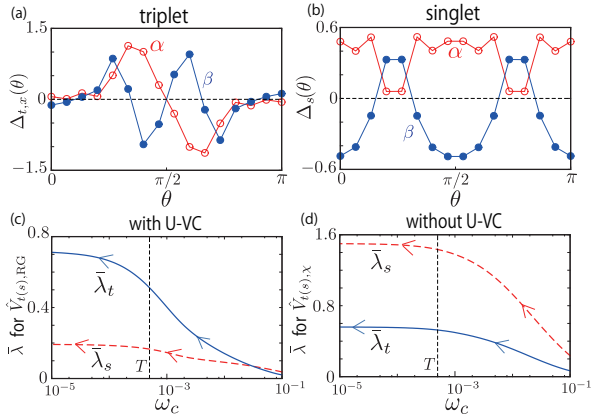}
\caption{(Color online)
(a) $E_{1u}$-type TSC Gap function $\Delta_{t,x}(\theta)$ 
on the FS$\a$ and FS$\b$ as functions of $\theta$.
(b) $A_{1g}$-type SSC Gap function $\Delta_{s}(\theta)$.
(c) $\bar{\lambda}_{t(s)}$ for ${\hat V}_{t(s),{\rm RG}}$
as functions of $\w_c$.
(d) $\bar{\lambda}_{t(s)}$ for ${\hat V}_{t(s),{\chi}}$.
}
\label{fig:gap}
\end{figure}

Hereafter, we show the numerical results for 
the parameters ($U=3.10$, $J/U=0.08$, $\w_c=12T=6\times 10^{-3}$),
which corresponds to the black circle in 
the phase diagram in Fig. \ref{fig:phase} (c).
The renormalization of ${\hat \chi}^{s,c}(\q)$ saturates for $\Lambda_l<\w_c$.
First, we solve the  gap equation (\ref{eqn:gap-eq})
using the pairing interaction ${\hat V}_{t,{\rm RG}}$ and
${\hat V}_{s,{\rm RG}}$ in Eqs. (\ref{eqn:V1t})-(\ref{eqn:V1s}).
Figures \ref{fig:gap} (a) and (b) show the 
obtained gap functions for the TSC state $\Delta_{t,x}(\theta)$
and the SSC state $\Delta_{s}(\theta)$, respectively,
The eigenvalues are $\lambda_t=0.47$ and $\lambda_s=0.26$, respectively.
The obtained $E_{1u}$ TSC gap and $A_{1g}$ SSC gap are 
essentially equivalent to the gap structures
derived from the SC susceptibilities in Eq. (\ref{eqn:suscepSC})
by the RG+cRPA: see Ref. \cite{Tsuchiizu2}.
Thus, the present gap equation analysis is essentially 
equivalent to the RG study for the SC state,
in which the SC gap function is uniquely obtained 
by maximizing the SC susceptibility.

Using the solution of the gap equation $\Delta_{t(s)}(\k)$,
the averaged pairing interaction
$\bar{\lambda}_{t(s)}={\lambda}_{t(s)}/{\rm ln}(1.13\w_c/T)$
is expressed as
\begin{eqnarray}
 \bar{\lambda}_{t(s)}
 = \frac{\displaystyle
 \int_{{\rm FS}} \frac{d\bm{k}}{v_{\bm{k}}}
 \int_{{\rm FS}} \frac{d\bm{k}'}{v_{\bm{k'}}}
  V_{t(s)}^{\w_c}({\k,\k'}) \Delta_{t(s)}({\bm k}) \Delta_{t(s)}({\bm k'})
  }
  {\displaystyle
 \int_{{\rm FS}} \frac{d{\bm k}}{v_{\bm k}} \Delta_{t(s)}({\bm k}) 
\Delta_{t(s)}({\bm k})} .
 \label{eqn:averaged}
\end{eqnarray} 
Figure \ref{fig:gap} (c) shows the obtained
$\bar{\lambda}_{t}$ and $\bar{\lambda}_{s}$
as functions of $\Lambda_l$,
where $\Delta_{t}(\k)$ and $\Delta_{s}(\k)$ are fixed to the 
gap structures shown in Figs. \ref{fig:gap} (a) and (b), respectively.
Note that the relation $T_{{\rm c},t(s)}=1.13\w_c\exp(-1/\bar{\lambda}_{t(s)})$.
The scaling curve of $\bar{\lambda}_{t,s}$ saturates 
to a constant when $\Lambda_l$ is smaller than $T$, 
which is shown by the vertical dotted lines.
We find the approximate relation $\bar{\lambda}_{t} \sim 3\bar{\lambda}_{s}$
in Fig. \ref{fig:gap} (c), irrespective of the relation 
$\chi^s(\Q)\sim\chi^c_{x^2-y^2}(\Q)$ shown in Figs. \ref{fig:phase} (a) and (b).

In order to verify the importance of the $U$-VC,
we solve the gap equation by using ${\hat V}_{x,\chi}$,
in which the $U$-VC is absent.
Figure \ref{fig:gap} (d) shows the obtained
$\bar{\lambda}_{t}$ and $\bar{\lambda}_{s}$
as functions of $\Lambda_l$.
Here, $\Delta_{t}(\k)$ and $\Delta_{s}(\k)$ are fixed to
Figs. \ref{fig:gap} (a) and (b), respectively.
(Similar result is obtained even if 
the solution of the gap equation for ${\hat V}_{t(s),\chi}$ is used.)
Thus, the relation $\bar{\lambda}_{t} \sim \bar{\lambda}_{s}/3$
is obtained if the $U$-VC is dropped.

Therefore,
the relation $\bar{\lambda}_{t} \gg \bar{\lambda}_{s}$ is realized
when $\hat{V}^{}_{t(s),{\rm RG}}$ is used, 
while the opposite relation $\bar{\lambda}_{t} \ll \bar{\lambda}_{s}$ 
is obtained for $\hat{V}^{}_{t(s),\chi}$. 
Thus, we can concluded that the TSC is realized by 
the enhancement of the orbital-fluctuation-mediated pairing 
interaction by the charge-channel $U$-VC,
and/or the suppression of the spin-fluctuation-mediated pairing 
by the spin-channel $U$-VC.

\begin{figure}[htb]
\includegraphics[width=.9\linewidth]{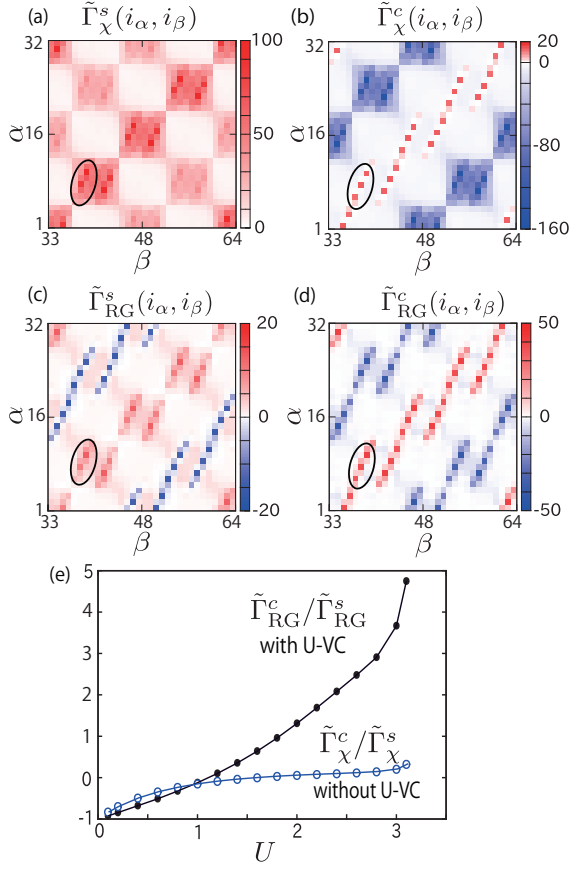}
\caption{(Color online)
Spin- and charge-channel pairing interactions
obtained by using the RG+cRPA method:
(a) Spin-channel interaction ${\tilde \Gamma}^s_\chi(\k,\k')$
and (b) charge-channel one ${\tilde \Gamma}^c_\chi(\k,\k')$
in the absence of the $U$-VC.
(c) ${\tilde \Gamma}^s_{\rm RG}(\k,\k')$
and (d) ${\tilde \Gamma}^c_{\rm RG}(\k,\k')$
in the presence of the $U$-VC.
Here, ($\k,\k'$) is the pair of momenta for ($i_\a,i_\b$).
(e) The ratios
${\tilde \Gamma}^c_\chi(\k,\k')/{\tilde \Gamma}^s_\chi(\k,\k')$ 
and  ${\tilde \Gamma}^c_{\rm RG}(\k,\k')/{\tilde \Gamma}^s_{\rm RG}(\k,\k')$
as functions of $U$.
$\k$ and $\k'$ are set as the start and end positions
of the nesting vector shown in Fig. \ref{fig:FS} (b).
We take the average over the ellipsoidal area.
}
\label{fig:interaction}
\end{figure}

To understand the role of the $U$-VC in more detail,
we directly examine the momentum-dependence of 
the spin- (charge-) channel interaction without the $U$-VC
${\tilde \Gamma}^{s(c)}_\chi(\k,\k') \equiv \Gamma^{s(c)}_\chi(\k,\k';-\k',-\k)$ 
in addition to those with the $U$-VC
${\tilde \Gamma}^{s(c)}_{\rm RG}(\k,\k') \equiv 
\Gamma^{s(c)}_{\rm RG}(\k,\k';-\k',-\k)$.
Figures \ref{fig:interaction} (a)-(d)
show the obtained interactions for 
the parameters ($U=3.10$, $J/U=0.08$, $\w_c=12T=6\times 10^{-3}$).
Here, $i_\a$ and $i_\b$ correspond to the patches 
on FS-$\a$ and FS-$\b$, respectively.
In each panel, the pairs of patches inside the solid ellipsoidal, 
$(i_\a,i_\b)=(6,37),(8,38),(10,39)$, correspond to the nesting vector 
$\k\rightarrow \k'$ depicted by the arrows in Fig. \ref{fig:FS} (c).

As shown in Figs. \ref{fig:interaction} (a) and (b),
both ${\tilde \Gamma}^{s}_\chi(\k,\k')$ and ${\tilde \Gamma}^{c}_\chi(\k,\k')$
take large positive values when
($i_\a,i_\b$) is inside the solid ellipsoidal.
Here, $\k-\k'\approx \Q \equiv (2\pi/3,2\pi/3)$.
These large interactions originates from the peak structure of
$\chi^{s}(\q)$ and $\chi^c_{x^2-y^2}(\q)$
at $\q\approx \Q$, as shown in Figs. \ref{fig:phase} (a) and (b).
It is found that, in the absence of the $U$-VC,
${\tilde \Gamma}^{s}_\chi(\k,\k')$ 
becomes larger than
${\tilde \Gamma}^{c}_\chi(\k,\k')$ inside the 
ellipsoidal area [$(i_\a,i_\b) \approx (7,37)$]
in Figs. \ref{fig:interaction} (a) and (b).
For this reason, the relation ${\bar \lambda}_s \gg {\bar \lambda}_t$
is realized by neglecting the $U$-VC, shown in Fig. \ref{fig:gap} (d).

Figures \ref{fig:interaction} (c) and (d) show
the spin- and charge-channel interactions
${\tilde \Gamma}^{s}_{\rm RG}(\k,\k')$ and ${\tilde \Gamma}^{c}_{\rm RG}(\k,\k')$
in the presence of the $U$-VC.
Both ${\tilde \Gamma}^{s}_{\rm RG}(\k,\k')$ and 
${\tilde \Gamma}^{c}_{\rm RG}(\k,\k')$
take large positive values when $\k-\k'\approx\Q$.
In the presence of the $U$-VC, 
${\tilde \Gamma}^{c}_{\rm RG}(\k,\k')$ 
becomes larger than
${\tilde \Gamma}^{s}_{\rm RG}(\k,\k')$ 
inside the ellipsoidal area.
By making comparison between 
Figs. \ref{fig:interaction} (a) and (c) [(b) and (d)],
the spin-channel [charge-channel] interaction 
is reduced [enlarged] by the $U$-VC.
For this reason, ${\bar \lambda}_t \gg {\bar \lambda}_s$
is realized by taking the $U$-VC into account correctly,
shown in Fig. \ref{fig:gap} (c).

We note that the large negative values
in Figs. \ref{fig:interaction} (c) and (d) 
at $(i_\a,i_\b)=(6+16,37),(8+16,38),(10+16,39)$
originate from ${\hat \chi}^c(\k+\k')$ for $\k+\k'\approx\Q$,
since its contribution is enlarged by the charge-channel $U$-VC
in ${\tilde \Gamma}^{s,c}_{\chi}(\k,\k')$.

Figure \ref{fig:interaction} (e) shows the ratios
${\tilde \Gamma}^c_\chi(\k,\k')/{\tilde \Gamma}^s_\chi(\k,\k')$ and  
${\tilde \Gamma}^c_{\rm RG}(\k,\k')/{\tilde \Gamma}^s_{\rm RG}(\k,\k')$
at $(i_\a,i_\b) \approx (8,38)$ [$\k-\k'\approx\Q$]
given by the RG+cRPA as functions of $U$.
We set $\w_c=12T=6\times10^{-3}$ and $J/U=0.08$.
$\k$ and $\k'$ are set as the start and end positions
of the nesting vector shown in Fig. \ref{fig:FS} (c).
For $U\rightarrow+0$, 
both ${\tilde \Gamma}^c_\chi/{\tilde \Gamma}^s_\chi$ and  
${\tilde \Gamma}^c_{\rm RG}/{\tilde \Gamma}^s_{\rm RG}$ are equal to $-1$.
They change to positive for $U \gtrsim 1$
since ${\tilde \Gamma}^c_{\chi({\rm RG})}$ changes to positive.
For $U\gtrsim2$, ${\tilde \Gamma}^c_\chi/{\tilde \Gamma}^s_\chi \ll1$ 
whereas ${\tilde \Gamma}^c_{\rm RG}/{\tilde \Gamma}^s_{\rm RG}\gg1$.
This result means that ${\tilde \Gamma}^{c(s)}_{\rm RG}$
is enlarged (suppressed) by the $U$-VC for wide range of $U$.

To summarize, 
the spin-channel [charge-channel] interaction 
is drastically reduced [enlarged] by the $U$-VC,
by making comparison between 
Figs. \ref{fig:interaction} (a) and (c) [(b) and (d)].
We stress that, except for the magnitude, the structure of 
${\tilde \Gamma}^{x}_{\rm RG}(\k,\k')$ and that of
${\tilde \Gamma}^{x}_{\chi}(\k,\k')$ ($x=s,c$)
are very similar.
In addition, when $\k$ and $\k'$ are on the same FS,
both ${\tilde \Gamma}^{x}_{\rm RG}$ 
and ${\tilde \Gamma}^{x}_\chi$ remain small.
These facts reveal the importance of the single-fluctuation-exchange term 
in Fig. \ref{fig:diagram} (b), since the 
multi-fluctuation-exchange terms such as in 
Fig. \ref{fig:diagram} (c) give different momentum dependence.
On the basis of the Fermi liquid theory,
the same charge-channel $U$-VC enlarges the charge
irreducible susceptibility ${\hat \Phi}^c(q)$ and
the pairing interaction, as we show in Fig. \ref{fig:fig1}.
Thus, the orbital-fluctuation-mediated pairing
will be strongly magnified by the $U$-VC
when the orbital fluctuations are driven by the VC.


\subsection{Analysis of the $U$-VC based on the perturbation theory}
\label{sec:UVC2}

In the previous section, we found the significant role of the
$U$-VC on the pairing interaction.
The orbital-fluctuation-mediated pairing interaction 
is strongly magnified by the charge channel $U$-VC.
We also found the strong suppression of 
the spin-fluctuation-mediated interaction
due to the spin-channel VC in multiorbital systems.
In this section, we perform the diagrammatic calculation for the $U$-VC
shown in Fig. \ref{fig:diagram} (e),
and confirm that the charge channel $U$-VC 
is strongly enlarged by the AL-VC.
In addition, the suppression by the spin channel $U$-VC
is mainly given by the $(U^0)^3$-term.
The charge- and spin-channel MT-terms in Fig. \ref{fig:diagram} (e)
are expressed as
\begin{eqnarray}
U^{c, {\rm MT}}_{l'm'lm} (k,k')
&=& \frac{T}{2} \sum_{q} \sum_{abcd} U^{0;c}_{l'm'bc} 
\big\{ I^{c}_{aldm}(q)+3 I^{s}_{aldm}(q) \big\} 
\nonumber \\
& &\times G_{ab}(k+q)G_{cd} (k'+q),
 \\
U^{s, {\rm MT}}_{l'm'lm} (k, k')
&=& \frac{T}{2} \sum_{q} \sum_{abcd} U^{0;s}_{l'm'bc} 
\big\{ I^{c}_{aldm}(q) - I^{s}_{aldm}(q) \big\} 
\nonumber \\
& &\times G_{ab} (k+q) G_{cd} (k'+q),
\end{eqnarray}
where
${\hat I}^x(q)= {\hat U}^{0;x}
({\hat \chi}^x_{\rm RPA}(q)+\{{\hat U}^{0;x}\}^{-1}){\hat U}^{0;x}$. 
Also, the charge- and spin-channel AL-terms in Fig. \ref{fig:diagram} (e) are
\begin{eqnarray}
&&U^{c, {\rm AL}}_{l'm'lm} (k, k')
=	\frac{T}{2} \sum_{q} \sum_{abcdefgh} U^{0;c}_{l'm'af} \nonumber \\
&& \times \big\{ \Lambda_{abcdef} (k - k', q) + \Lambda_{fcbeda} (k - k', - q - k + k') \big\} \nonumber \\
&& \times \big\{ I^{c}_{bclg} (q + k - k') I^{c}_{mhed} (q) + 3 I^{s}_{bclg} (q + k - k') I^{s}_{mhed} (q) \big\} \nonumber \\
&& \times G_{gh} (k' - q) ,
\label{eqn:ALc} \\
&&U^{s, {\rm AL}}_{l'm'lm} (k, k')
	= \frac{T}{2} \sum_{q} \sum_{abcdefgh} U^{0;s}_{l'm'af} \nonumber \\
	&& \times \big\{ \Lambda_{abcdef} (k - k', q) + \Lambda_{fcbeda} (k - k', - q - k + k') \big\} \nonumber \\
	&& \times \big\{ I^{s}_{bclg} (q + k - k') I^{c}_{mhed} (q) + I^{c}_{bclg} (q + k - k') )I^{s}_{mhed} (q) \big\} \nonumber \\
	&& \times G_{gh} (k' - q)
\nonumber \\
&& + \delta U^{s, {\rm AL}}_{l'm'lm} (k, k'),
\label{eqn:ALs}
\end{eqnarray}
where $a\sim h$ are orbital indices, and
${\hat \Lambda}(q,q')$ is the three-point vertex given as
\begin{eqnarray}
\Lambda_{abcdef} (q, q') = - T \sum_{p} G_{ab} (p + q) G_{cd} (p - q') G_{ef} (p) .
\end{eqnarray}
The last term in Eq. (\ref{eqn:ALs}) is given as
$\delta U^{s, {\rm AL}}_{l'm'lm} (k, k')
=\frac{T}{2} \sum_{q} \sum_{abcdefgh} U^{s, 0}_{l'm'af} 
\big\{ \Lambda_{abcdef} (k - k', q) - 
\Lambda_{fcbeda} (k - k', - q - k + k') \big\} 
 2 I^{s}_{bclg} (q + k - k') I^{s}_{mhed} (q) G_{gh} (k' - q)$,
which is found to be very small.

\begin{figure}[htb]
\includegraphics[width=.9\linewidth]{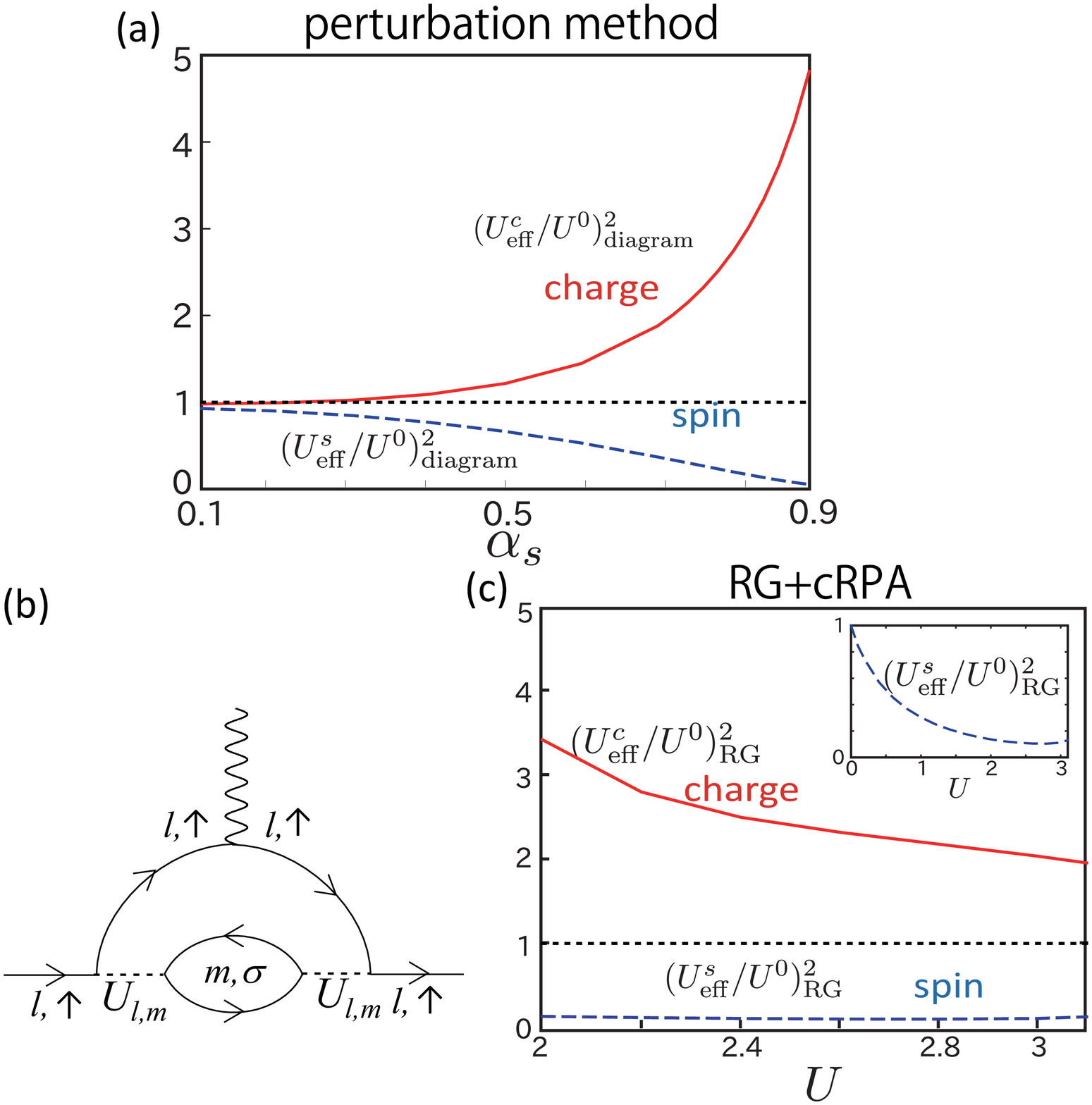}
\caption{(Color online)
(a) The ratios
$(U^x_{\rm eff}/U^0)^2_{\rm diagram} \equiv 
(U^x_{\rm with\mbox{-}{\it U}VC}(\k,\k')/U^x_{\rm no\mbox{-}{\it U}VC}(\k,\k'))^2$ ($x=c,s$)
given by the diagrammatic calculation as functions of the
spin Stoner factor $\a_S$.
For $U$-VC, we perform the diagrammatic calculation for 
Fig. \ref{fig:diagram} (e).
(b) Third-order term with respect to $U$ for $U$-VC:
We put $U=U'$ and $J=0$ for simplicity.
This term is scaled as $\sim (2N_{\rm orb}-1)$,
where $N_{\rm orb}$ is the number of $d$-orbital.
(c) $(U^x_{\rm eff}/U^0)^2_{\rm RG} \equiv
{\tilde \Gamma}^x_{\rm RG}/{\tilde \Gamma}^x_{\chi}$ 
given by the RG+cRPA method for $2.0\le U \le 3.1$.
Inset: $(U^s_{\rm eff}/U^0)^2_{\rm RG}$ for $0\le U\le 3.1$.
}
\label{fig:perturbation}
\end{figure}

Figure \ref{fig:perturbation} (a) shows the ratios
$(U^x_{\rm eff}/U^0)^2_{\rm diagram} \equiv 
(U^x_{\rm with\mbox{-}{\it U}VC}(\k,\k')/U^x_{\rm no\mbox{-}{\it U}VC}(\k,\k'))^2$ ($x=s,c$)
at $(i_\a,i_\b) \approx (8,38)$ [$\k-\k'\approx\Q$]
given by the diagrammatic calculation as functions of 
the spin Stoner factor $\a_S$.
For $U$-VC, we perform the diagrammatic calculation for 
Fig. \ref{fig:diagram} (e).
The double counting of the $O(\{U^0\}^3)$-terms
is carefully eliminated.
Note that $\a_S$ is the largest eigenvalue of 
${\hat \Gamma}^s{\hat\chi}^0(\Q)$,
and the relation $\chi^s(\Q)\propto (1-\a_S)^{-1}$ holds.
We find that $(U^c_{\rm eff}/U^0)^2_{\rm diagram}$ gradually increases as the system 
approaches to the magnetic quantum-critical-point ($\a_S\rightarrow1$).
The relation $(U^c_{\rm eff}/U^0)_{\rm diagram}^2\gg1$ originates from the 
charge-channel AL-term since Eq. (\ref{eqn:ALc})
is approximately proportional to 
$\sum_\q \chi^s(\q)\chi^s(\q+\Q) \sim (1-\a_S)^{-1}$.
In contrast, $(U^s_{\rm eff}/U^0)_{\rm diagram}^2$ is suppressed by the $U$-VC,
since the small spin-channel AL-term in Eq. (\ref{eqn:ALs}) 
is proportional to $\sum_\q \chi^s(\q)\chi^c(\q+\Q)$.
We verified that the relation $(U^s_{\rm eff}/U^0)^2_{\rm diagram} \ll1$
mainly originates from the $O(\{U^0\}^3)$-term shown in 
Fig. \ref{fig:perturbation} (b):
Its negative contribution is significant in multiorbital systems
since the diagram in Fig. \ref{fig:perturbation} (b) 
is scaled as $\sim(2N_{\rm orb}-1)$, where 
$N_{\rm orb}$ is the number of $d$-orbital.

Figure \ref{fig:perturbation} (c) shows
$(U^x_{\rm eff}/U^0)^2_{\rm RG} 
\equiv {\tilde \Gamma}^x_{\rm RG}(\k,\k')/{\tilde \Gamma}^x_{\chi}(\k,\k')$ 
($x=s,c$) at $(i_\a,i_\b) \approx (8,38)$ [$\k-\k'\approx\Q$]
obtained by the RG+cRPA study as function of $U$.
Here, $\w_c=12T=6\times10^{-3}$ and $J/U=0.08$.
This ratio is expect to give the square of the $U$-VC
when ${\hat \chi}^{s,c}(\q)$ develops strongly 
in the strong-coupling region ($U\gtrsim2.5$), 
in which the single-fluctuation-exchange term in 
Fig. \ref{fig:diagram} (b) becomes significant.
The obtained relations
$(U^c_{\rm eff}/U^0)_{\rm RG}^2\gg1$ and $(U^s_{\rm eff}/U^0)_{\rm RG}^2 \ll1$ 
in the strong-coupling region
are consistent with the results given by the 
perturbation theory in Fig. \ref{fig:perturbation} (a).
The inset shows $(U^s_{\rm eff}/U^0)^2_{\rm RG}$ for wide range of $U$:
The origin of its $U$-linear term for $U\sim0$ would be
some $U^2$-diagrams dropped in ${\tilde \Gamma}^x_{\chi}$,
which are less important for the strong-coupling region.
(Note that $(U^c_{\rm eff}/U^0)^2_{\rm RG}$ diverges at $U\approx 1.5$
since ${\tilde \Gamma}^x_{\chi}(\k,\k')$ changes its sign with $U$;
see in Fig. \ref{fig:interaction} (e).)

In summary, 
the significant role of the $U$-VC has been confirmed 
on the basis of the perturbation theory and the RG+cRPA theory.
Due to the $U$-VC,
the orbital- or charge-fluctuation-mediated pairing interaction
is magnified by $(U^c_{\rm eff}/U^0)^2\gg1$
in the strong-coupling regime.
In contrast, the spin-fluctuation-mediated pairing interaction
is suppressed by $(U^s_{\rm eff}/U^0)^2\ll1$,
and this suppression is prominent in multiorbital systems.
In the strong-coupling regime,
consistent results are obtained by the different two methods
shown in Figs. \ref{fig:perturbation} (a) and (c).
They do not coincide in the weak coupling regime
because of the different definitions of $(U^x_{\rm eff}/U^0)^2$
in Figs. \ref{fig:perturbation} (a) and (c).

\section{Discussions}
\label{sec:dis}

In this paper,
we analyzed the two-orbital Hubbard model by using the RG+cRPA theory
in order to confirm the realization condition for the 
orbital-fluctuation-mediated SC.
To go beyond the ME approximation,
we solved the gap equation by including the VC for the EBC, 
which is called the $U$-VC.
Due to the $U$-VC, the effective EBC for the charge (spin) channel,
${\hat U}^{c(s)}$, deviates from 
the bare Coulomb interaction ${\hat U}^{0;c(s)}$.
We verified the relation 
$|{\hat U}^{c}|\gg |{\hat U}^{0;c}|$ due to the charge-channel $U$-VC
in the presence of moderate spin fluctuations.
In contrast, 
${\hat U}^{s}$ is significantly suppressed by the spin channel $U$-VC.
For these reasons, 
orbital-fluctuation-mediated SC will be realized 
in various multiorbital systems, 
such as in Fe-based superconductors and Sr$_2$RuO$_4$.

On the basis of the Fermi liquid theory,
the same charge-channel $U$-VC enlarges the charge
irreducible susceptibility ${\hat \Phi}^c(q)$ and
the pairing interaction, as we show in Fig. \ref{fig:fig1}.
Thus, the orbital-fluctuation-mediated pairing interaction
should be strongly enlarged by the square of the $U$-VC
when the orbital fluctuations are driven by the VC
in terms of the Fermi liquid theory.

In fact, the importance of the single-fluctuation-exchange term 
in Fig. \ref{fig:diagram} (b) is
supported by the very similar momentum dependence 
between ${\tilde \Gamma}^{x}_{\rm RG}(\k,\k')$ and  
${\tilde \Gamma}^{x}_{\chi}(\k,\k')$ ($x=c,s$) 
in Fig. \ref{fig:interaction} (a)-(d),
except for the magnitude.
The drastic difference in magnitude between 
${\tilde \Gamma}^{x}_{\rm RG}$ and ${\tilde \Gamma}^{x}_{\chi}$ 
demonstrates the significance of the $U$-VC.
We verified that the 
crossing-fluctuation-exchange term in Fig. \ref{fig:diagram} (c),
which should have different momentum dependence,
is small in magnitude based on the perturbation method.

\begin{figure}[htb]
\includegraphics[width=.8\linewidth]{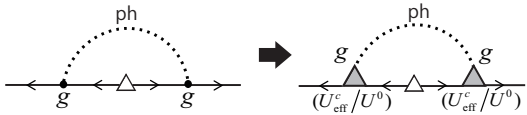}
\caption{The gap equation due to the $e$-ph interaction,
where the dotted line represents the phonon propagator
and $g$ is the $e$-ph coupling constant.
Due to the charge-channel $U$-VC 
caused by spin fluctuations,
the phonon-mediated attractive interaction is 
enlarged by the factor $(U^c_{\rm eff}/U^0)^2\gg1$.
}
\label{fig:phonon}
\end{figure}

We stress that the phonon-mediated attractive pairing 
is also enlarged by the factor $({U}_{\rm eff}^{c}/{U}^{0})^2\gg1$,
as we explain in Fig. \ref{fig:phonon}.
The $s_{++}$-wave state in the single-layer FeSe
may be given by the electron-phonon ($e$-ph) attractive interaction 
enhanced by the charge-channel $U$-VC.
Note that the relation $({U}_{\rm eff}^{c}/{U}^{0})^2\gg1$
in the presence of moderate spin fluctuations is realized 
only in two- and three-dimensional systems.
If we apply the local approximation,
the charge-channel VC is proportional to the square of $\sum_q\chi^s(q)$, 
which is less singular even for $\a_S\approx 1$.

In multiorbital models,
the spin-fluctuation-mediated pairing interaction 
is strongly suppressed by the factor $({U}_{\rm eff}^{s}/{U}^{0})^2\ll1$.
This result does not contradict to the enhancement of spin susceptibility 
$\chi^s(\q)$ shown in Fig. \ref{fig:diagram} (a),
since the $U$-VC is effective only at low energies,
whereas the irreducible susceptibility $\Phi^s$ in Fig. \ref{fig:fig1} (b)
is given by the integration for wide energy range.
In the context of the fRG,
$\chi^s(\q)$ starts to increase in the early stage of the renormalization,
whereas the $U$-VC develops in the later stage.

\acknowledgments
We are grateful to W. Metzner and C. Honerkamp
for useful comments and discussions. This study has been supported by Grants-in-Aid for Scientific Research from Ministry of Education, Culture, Sports, Science, and Technology of Japan.


\end{document}